\documentclass[runningheads]{llncs}
\usepackage[T1]{fontenc}
\usepackage{graphicx}
\usepackage[linesnumbered,vlined]{algorithm2e}
\usepackage{booktabs}
\usepackage{adjustbox}
\usepackage{amsmath,mathtools}
\usepackage{graphics}
\usepackage{multirow}
\usepackage{xcolor}
\usepackage{subfig}
\usepackage{graphicx}
\usepackage{hhline}
\usepackage{makecell}
\usepackage{hyperref}
\usepackage{arydshln}
\usepackage{colortbl}
\usepackage{enumitem}
\usepackage{relsize}
\usepackage{courier}
\usepackage{xspace}
\usepackage[utf8]{inputenc}
\usepackage[super]{nth}
\usepackage{dsfont}
\usepackage{xspace}
\usepackage{enumitem}

\SetKwComment{Comment}{/* }{ */}
\newcommand{\colbert}{ColBERT\xspace}
\newcommand{\colberttwo}{ColBERTv2\xspace}

\newcommand{\msmarco}{MS MARCO\xspace}

\DeclarePairedDelimiter\ceil{\lceil}{\rceil}

\begin{document}

\title{Efficient Multi-Vector Dense Retrieval\\with Bit Vectors}
\titlerunning{Efficient Multi-Vector Dense Retrieval with Bit Vectors}

\author{Franco Maria Nardini\inst{1} \and Cosimo Rulli\inst{1} \and Rossano Venturini\inst{2}}


\institute{ISTI-CNR, Pisa, Italy 
\email{\{name.surname\}@isti.cnr.it} \and
 University of Pisa, Italy
\email{rossano.venturini@unipi.it}}

\maketitle

\begin{abstract}
Dense retrieval techniques employ pre-trained large language models to build a high-dimensional representation of queries and passages. These representations compute the relevance of a passage w.r.t. to a query using efficient similarity measures. In this line, multi-vector representations show improved effectiveness at the expense of a one-order-of-magnitude increase in memory footprint and query latency by encoding queries and documents on a per-token level.
Recently, PLAID has tackled these problems by introducing a centroid-based term representation to reduce the memory impact of multi-vector systems. By exploiting a centroid interaction mechanism, PLAID filters out non-relevant documents, thus reducing the cost of the successive ranking stages.
This paper proposes ``Efficient Multi-Vector dense retrieval with Bit vectors'' (EMVB), a novel framework for efficient query processing in multi-vector dense retrieval. 
First, EMVB employs a highly efficient pre-filtering step of passages using optimized bit vectors. Second, the computation of the centroid interaction happens column-wise, exploiting SIMD instructions, thus reducing its latency. Third, EMVB leverages Product Quantization (PQ) to reduce the memory footprint of storing vector representations while jointly allowing for fast late interaction. Fourth, we introduce a per-document term filtering method that further improves the efficiency of the last step. 
Experiments on MS MARCO and LoTTE show that EMVB is up to $2.8\times$ faster while reducing the memory footprint by $1.8\times$ with no loss in retrieval accuracy compared to PLAID.


\end{abstract}


\section{Introduction}
The introduction of pre-trained large language models (LLM) has remarkably improved the effectiveness of information retrieval systems~\cite{khattab2021baleen,guu2020retrieval,zhan2021optimizing,DBLP:journals/ftir/BruchLN23}, thanks to the well-known ability of LLMs to model semantic and context~\cite{kenton2019bert,brown2020language,dai2022can}.
In dense retrieval, LLMs have been successfully exploited to learn high-dimensional dense representations of passages and queries. These learned representations allow answering the user query through fast similarity operations, i.e., inner product or L2 distance. 
In this line, multi-vector techniques~\cite{khattab2020colbert,santhanam2022colbertv2} employ an LLM to build a dense representation for each token of a passage. These approaches offer superior effectiveness compared to single-vector techniques~\cite{xiongapproximate,zhan2022learning} or sparse retrieval techniques~\cite{formal2021splade}. 
In this context, the similarity between the query and the passage is measured by using the \emph{late interaction} mechanism~\cite{khattab2020colbert,santhanam2022colbertv2}, which works by computing the sum of the maximum similarities between each term of the query and each term of a candidate passage. 
The improved effectiveness of multi-vector retrieval system comes at the price of its increased computational burden. First, producing a vector for each token causes the number of embeddings to be orders of magnitude larger than in a single-vector representation. Moreover, due to the large number of embeddings, identifying the candidate documents\footnote{the terms ``document'' and ``passage'' are used interchangeably in this paper.} is time-consuming. In addition, the late interaction step requires computing the maximum similarity operator between all the candidate embeddings and the query, which is also time-consuming.

Early multi-vector retrieval systems, i.e., \colbert~\cite{khattab2020colbert}, exploit an inverted index to store the embeddings and retrieve the candidate passages. Then, the representations of passages are retrieved and employed to compute the max-similarity score with the query. Despite being quite efficient, this approach requires maintaining the full-precision representation of each document term in memory. On \msmarco~\cite{nguyenms}, a widely adopted benchmark dataset for passage retrieval, the entire collection 
of embeddings used by \colbert requires more than $140$ GB ~\cite{khattab2020colbert} to be stored.
\colberttwo~\cite{santhanam2022colbertv2} introduces a centroid-based compression technique to store the passage embeddings efficiently. Each embedding is stored by saving the $id$ of the closest centroid and then compressing the residual (i.e., the element-wise difference) by using $1$ or $2$ bits per component. 
\colberttwo saves up to $10\times$ space compared to \colbert while being significantly more inefficient on modern CPUs, requiring up to $3$ seconds to perform query processing on CPU~\cite{santhanam2022plaid}.
The reduction of query processing time is achieved by Santhanam \emph{et al.} with PLAID~\cite{santhanam2022plaid}. PLAID takes advantage of the embedding compressor of \colberttwo and also uses the centroid-based representation to discard non-relevant passages (\emph{centroid interaction}~\cite{santhanam2022plaid}), thus performing the late interaction exclusively on a carefully selected batch of passages. PLAID allows for massive speedup compared to \colberttwo, but its average query latency can be up to $400$ msec. on CPU with single-thread execution~\cite{santhanam2022plaid}.


This paper presents EMVB, a novel framework for efficient query processing with multi-vector dense retrieval. First, we identify the most time-consuming steps of PLAID. These steps are i) extracting the top-$nprobe$ closest centroids for each query term during the candidate passage selection, ii) computing the centroid interaction mechanism, and iii) decompression of the quantized residuals. Our method tackles the first and the second steps by introducing a highly efficient passage filtering approach based on optimized bit vectors. Our filter identifies a small set of crucial centroid scores, thus tearing down the cost of top-$nprobe$ extraction. At the same time, it reduces the amount of passages for which we have to compute the centroid interaction. 
Moreover, we introduce a highly efficient column-wise reduction exploiting SIMD instructions to speed up this step. 
Finally, we improve the efficiency of the late interaction by introducing Product Quantization (PQ)~\cite{jegou2010product}. PQ allows to obtain in pair or superior performance compared to the bitwise compressor of PLAID while being up to $3\times$ faster. Finally, to further improve the efficiency of the last step of our pipeline, we introduce a dynamic passage-term-selection criterion for late interaction, thus reducing the cost of this step up to $30\%$.

We experimentally evaluate EMVB against PLAID on two datasets: \msmarco passage~\cite{nguyenms} (for in-domain evaluation) and LoTTE~\cite{santhanam2022colbertv2} (for out-of-domain evaluation). Results on \msmarco show that EMVB is up to $2.8\times$ faster while reducing the memory footprint by $1.8\times$ with no loss in retrieval accuracy compared to PLAID. On the out-of-domain evaluation, EMVB delivers up to $2.9\times$ speedup compared to PLAID, with a minimal loss in retrieval quality.

The rest of this paper is organized as follows. In Section~\ref{sec:relwork}, we discuss the related work. In Section~\ref{sec:mt_dense_retr} we describe PLAID~\cite{santhanam2022plaid}, the current state-of-the-art in multi-vector dense retrieval. We introduce EMVB in Section~\ref{sec:meth} and we experimentally evaluate it against PLAID in Section~\ref{sec:exp}. Finally, Section~\ref{sec:conclusion} concludes our work.

\section{Related Work}
\label{sec:relwork}
Dense retrieval encoders can be broadly classified into single-vector and multi-vector techniques. Single-vector encoders allow the encoding of an entire passage in a single dense vector \cite{karpukhin2020dense}. In this line, ANCE~\cite{xiong2020approximate} and STAR/ADORE~\cite{zhan2021optimizing} employ hard negatives to improve the training of dense retrievers by teaching them to distinguish between lexically-similar positive and negative passages. Multi-vector encoders have been introduced with \colbert. The limitations of \colbert and the efforts done to overcome them (\colberttwo, PLAID) have been discussed in Section 1.
COIL~\cite{gao2021coil} rediscover the lessons of classical retrieval systems (e.g., BM25) by limiting the token interactions to lexical matching between queries and documents.
CITADEL~\cite{li2022citadel} is a recently proposed approach that introduces conditional token interaction by using dynamic lexical routing. Conditional token interaction means that the relevance of the query of a specific passage is estimated by only looking at some of their tokens. These tokens are selected by the so-called lexical routing, where a module of the ranking architecture is trained to determine which of the keys, i.e., words in the vocabulary, are activated by a query/passage. CITADEL significantly reduces the execution time on GPU, but turns out to be $2\times$ slower than PLAID con CPU, at the same retrieval quality.
%
Multi-vector dense retrieval is also exploited in conjunction with pseudo-relevance feedback both in ColBERT-PRF~\cite{wang2023colbert} and in CWPRF~\cite{wang2023effective}, showing that their combination boosts the effectiveness of the model.

\vspace{1mm}
\noindent \textbf{Our Contribution}: This work advances the state of the art of multi-vector dense retrieval by introducing EMVB, a novel framework that allows to speed up the retrieval performance of the PLAID pipeline significantly. To the best of our knowledge, this work is the first in the literature that proposes a highly efficient document filtering approach based on optimized bit vectors, a column-wise SIMD reduction to retrieve candidate passages and a late interaction mechanism that combines product quantization with a per-document term filtering.


\section{Multi-vector Dense Retrieval}
\label{sec:mt_dense_retr}
Consider a passage corpus $\mathcal{P}$ with $n_P$ passages. In a multi-vector dense retrieval scenario, an LLM encodes each token in $\mathcal{P}$ into a dense $d$-dimensional vector $T_j$. For each passage $P$, a dense representation $P = \{ T_j \}$, with $j=0, \dots, n_t$, is produced, where $n_t$ is the number of tokens in the passage $P$.
Employing a token-level dense representation allows for boosting the effectiveness of the retrieval systems~\cite{khattab2020colbert,santhanam2022colbertv2,santhanam2022plaid}. 
On the other hand, it produces significantly large collections of $d$-dimensional vectors posing challenges to the applicability of such systems in real-world search scenarios both in terms of space (memory requirements) and time (latency of the query processor).
To tackle the problem of memory requirements, \colberttwo~\cite{santhanam2022colbertv2} and successively PLAID~\cite{santhanam2022plaid} exploit a centroid-based vector compression technique. 
First, the K-means algorithm is employed to devise a clustering of the $d$-dimensional space by identifying the set of $k$ centroids $\mathcal{C} = \{C_i\}_{i=1}^{n_c}$.
Then, for each vector $x$, the residual $r$ between $x$ and its closest centroid $\bar{C}$ is computed so that $r = x - \bar{C}$.
The residual $r$ is compressed into $\tilde{r}$ using a $b$-bit encoder that represents each dimension of $r$ using $b$ bits, with $b \in \{1, 2\}$.
The memory impact of storing a $d$-dimensional vector is given by $\ceil*{\log_2{|C|}}$ bits for the centroid index and $d \times b$ bits for the compressed residual.
This approach requires a time-expensive decompression phase to restore the approximate full-precision vector representation given the centroid id and the residual coding. 
For this reason, PLAID aims at decompressing as few candidate documents as possible. This is achieved by introducing a high-quality filtering step based on the centroid-approximated embedding representation, named \emph{centroid interaction}~\cite{santhanam2022plaid}. 
In detail, the PLAID retrieval engine is composed of four different phases ~\cite{santhanam2022plaid}. 
The first one regards the \emph{retrieval} of the candidate passages. 
A list of candidate passages is built for each centroid.
A passage belongs to a centroid $C_i$ candidate list if one or more tokens have $C_i$ as its closest centroid. For each query term $q_i$, with $i= 1,\dots, n_q$, the top-\emph{nprobe} closest centroids are computed, according to the \emph{dot product} similarity measure.
The set of unique documents associated with the top-\emph{nprobe} centroids then moves to a second phase that acts as a \emph{filtering} phase. In this phase, a token embedding $T_j$ with $j=1, \dots, n_p$ is approximated using its closest centroid $\bar{C}^{T_j}$. Hence, its distance with the $i$-th query term $q_i$ is approximated with
\begin{equation}
    \label{eq:tj}
    q_i \cdot T_j \simeq q_i \cdot \bar{C}^{T_j} = \tilde{T}_{i,j}.
\end{equation} 
Consider a candidate passage $P$ composed of $ n_p$ tokens. The approximated score of $P$ consists in computing the dot product $q_i \cdot \bar{C}^{T_j}$ for all the query terms $q_i$ and all the closest centroids of each token belonging to the passage, i.e.,
\begin{equation}
\label{eq:plaid_approx}
\bar{S}_{q, P} = \sum_{i=1}^{n_q} \max_{j=1 \dots n_t} q_i \cdot \bar{C}^{T_j}
\end{equation}
The third phase, named \emph{decompression}, aims at reconstructing the full-precision representation of $P$ by combining the centroids and the residuals. This is done on the top-\emph{ndocs} passages selected according to the \emph{filtering} phase~\cite{santhanam2022plaid}. In the fourth phase, PLAID recomputes the final score of each passage with respect to the query $q$ using the decompressed---full-precision---representation according to \emph{late interaction} mechanism (Equation~\ref{eq:maxsim}). Passages are then ranked according to their similarity score and the top-$k$ passages are selected.
\begin{equation}
\label{eq:maxsim}    
S_{q, P} = \sum_{i=1}^{n_q} \max_{j =1\dots n_t} q_i \cdot T_j.
\end{equation}

\noindent \textbf{PLAID execution time}. 
We provide a breakdown of PLAID execution time across its different phases, namely \emph{retrieval}, \emph{filtering}, \emph{decompression}, and \emph{late interaction}.
This experiment is conducted using the experimental settings detailed in Section \ref{sec:exp}.
We report the execution time for different values of $k$, i.e., the number of retrieved passages. 

\begin{figure}[h!]
\centering
\includegraphics[width=\columnwidth]{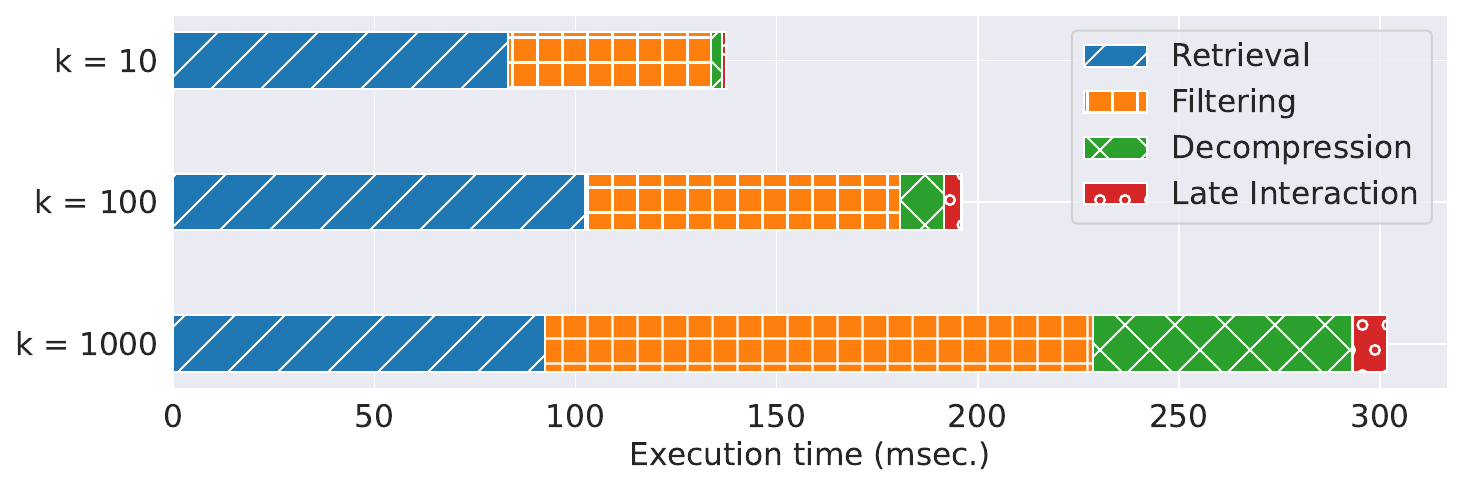}
\caption{Breakdown of the PLAID average query latency (in milliseconds)  on CPU across its four phases.\label{fig:plaid_breakdown}}
\end{figure}

\section{EMVB}
\label{sec:meth}
We now present EMVB, our novel framework for efficient multi-vector dense retrieval. First, EMVB introduces a highly efficient pre-filtering phase that exploits optimized bit vectors. Second, we improve the efficiency of the centroid interaction step (Equation 1) by introducing column-wise max reduction with SIMD instructions. Third, EMVB leverages Product Quantization (PQ) to reduce the memory footprint of storing the vector representations while jointly allowing for a fast late interaction phase. Fourth, PQ is applied in conjunction with a novel per-passage term filtering approach that allows for further improving the efficiency of the late interaction. In the following subsections, we detail these four contributions behind EMVB.

\subsection{Retrieval of Candidate Passages}
\label{subsec:centr_ident}
Figure~\ref{fig:plaid_breakdown} shows that a consistent part of the computation required by PLAID is spent on the retrieval phase. We further break down these steps to evidence its most time-consuming part. The retrieval
consists of i) computing the distances between the incoming query and the set of centroids, ii) extracting the top-$nprobe$ closest centroids for each query term. 
The former step is efficiently carried out by leveraging high-performance matrix multiplication tools (e.g.,  Intel MKL~\cite{qian2004similarity,wang2014intel}). 
In the latter step, PLAID extracts the top-$nprobe$ centroids using the numpy \texttt{topk} function, which implements the \emph{quickselect} algorithm. Selecting the top-$nprobe$ within the $|C| = 2^{18}$ centroids for each of the $n_q$ query terms costs up to $3\times$ the matrix multiplication done in the first step.
In Section~\ref{subsec:docselection}, we show that our pre-filtering inherently speeds up the top-$nprobe$ selection by tearing down the number of evaluated elements. In practice, we show how to efficiently filter out those centroids whose score is below a certain threshold and then execute quickselect exclusively on the surviving ones. 
As a consequence, in EMVB the cost of the top-$nprobe$ extraction becomes negligible, being two orders of magnitude faster than the  top-$nprobe$ extraction on the full set of centroids.

\subsection{Efficient Pre-Filtering of Candidate Passages}
\label{subsec:docselection}
Figure~\ref{fig:plaid_breakdown} shows that the candidate filtering phase can be significantly expensive, especially for large values of $k$. 
In this section, we propose a pre-filtering approach based on a novel bit vector representation of the centroids that efficiently allows the discarding of non-relevant passages. 

Given a passage $P$, our pre-filtering consists in determining whether $\tilde{T}_{i,j}$, for $i=1, \dots, n_q$, $j =1, \dots, n_t$  is large or not. 
Recall that $\tilde{T}_{i,j}$ represents the approximate score of the $j$-th token of passage $P$ with respect to the $i$-th term of the query $q_i$, as defined in Equation~\ref{eq:tj}. 
This can be obtained by checking whether $\bar{C}^{T}_j$---the centroid associated with $T_j$---belongs to the set of the \emph{closest centroids} of $q_i$.
We introduce $\mathtt{close}^{th}_i$, the set of centroids whose scores are greater than a certain threshold $th$ with respect to a query term $q_i$. 
Given a passage $P$,
we define the list of centroids ids $I_P$, where $I_P^j$ is the centroid id of $\bar{C}^{T_j}$.
The similarity of a passage with respect to a query can be estimated with our novel filtering function $F(P,q)\in [0, n_q]$ with the following equation:
\begin{equation}
\label{eq:our_filter_th}
F(P, q) =  \sum_{i=1}^{n_q} \mathbf{1}(\exists \  j \text{ s.t. } I_P^j \in \mathtt{close}^{th}_i).
\end{equation}
For a passage $P$, this counts how many query terms have at least one similar passage term in $P$, where ``similar'' describes the belonging of $T_j$ to $\mathtt{close}^{th}_i$. 

\begin{figure}[!tb]
\centering
\includegraphics[width=\columnwidth]{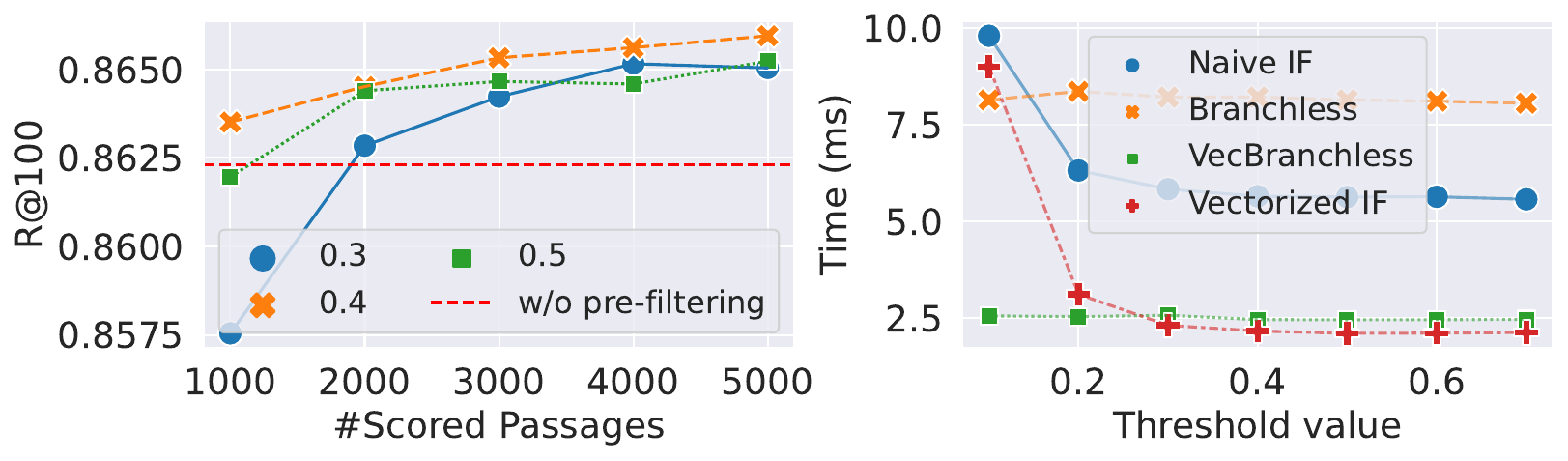}
\caption{R@100 with various values of the threshold (\textbf{left}). Comparison of different algorithms to construct $\mathtt{close}^{th}_i$, for different values of $th$ (\textbf{right}).\label{fig:first_mixed}}
\end{figure}

In Figure~\ref{fig:first_mixed} (left), we compare the performance of our novel pre-filter working on top of the centroid interaction mechanism (orange, blue, green lines) against the performance of the centroid interaction mechanism on the entire set of candidate documents (red dashed line) on the \msmarco dataset. The plot shows that our pre-filtering allows to efficiently discard non-relevant passages without harming the recall of the successive centroid interaction phase. For example, we can narrow the candidate passage set to just $1000$ elements using $th=0.4$ without any loss in R@100. 
In the remainder of this section, we show how to implement this pre-filter efficiently. 

\vspace{1mm}
\noindent \textbf{Building the bit vectors}.
Given $th$, the problem of computing $\mathtt{close}^{th}_i$ 
is conceptually simple. Yet, an efficient implementation carefully considering modern CPUs' features is crucial for fast computation of Equation~\ref{eq:our_filter_th}.

Let $CS = q \cdot C^{T}$, with $CS \in [-1, 1]^{n_q \times |C|}$ be the score matrix between the query $q$ and the set of centroids $C$ (both matrices are $L_2$ normalized), where $n_q$ is the number of query tokens, and $|C|$ is the number of centroids.
In the na\"ive \emph{if}-based solution, we scan the $i$-th row of $CS$ and select those $j$ s.t. $CS_{i,j} > th$.
It is possible to speed up this approach by taking advantage of SIMD instructions. 
In particular, the \texttt{\_mm512\_cmp\_epi32\_mask} allows one to compare $16$ fp32 values at a time and store the comparison result in a $mask$ variable. If $mask==0$, we can skip to the successive $16$ values because the comparison has failed for all the current $j$s. Otherwise, we extract those indexes $J = \{ j \in [0, 15] \ | \ mask_j = 1 \}$. 

The efficiency of such \emph{if}-based algorithms mainly depends on
the \emph{branch misprediction} ratio. Modern CPUs speculate on the outcome of the \emph{if} before the condition itself is computed by recognizing patterns in the execution flow of the algorithm. When the wrong branch is predicted, a \emph{control hazard} happens, and the pipeline is flushed with a delay of $15-20$ clock cycles, i.e., about $10$~ns. 
We tackle the inefficiency of branch misprediction by proposing a \emph{branchless} algorithm. 
The branchless algorithm employs a pointer $p$ addressing a pre-allocated buffer. While scanning $CS_{i,j}$, it writes $j$ in the position indicated by $p$. Then, it sums to $p$ the result of the comparison: $1$ if $CS_{i,j} > th$, $0$ otherwise. At the successive iteration, if the result of the comparison was $0$, $j+1$ will override $j$. Otherwise, it will be written in the successive memory location, and $j$ will be saved in the buffer. 
The branchless selection does not present any \emph{if} instruction and consequently does not contain any branch in its execution flow. 
The branchless algorithm can be implemented more efficiently by leveraging SIMD instructions. In particular, the above-mentioned  \texttt{\_mm512\_cmp\_epi32\_mask}  instruction allows to compare $16$ fp32 values at the time, and the \texttt{\_mm512\_mask\_compressstore} allows to extract $J$ in a single instruction. 

Figure~\ref{fig:first_mixed} (right) presents a comparison of our different approaches, namely ``Na\"ive IF'', the ``Vectorized IF'', the ``Branchless'', and the ``VecBranchless'' described above. 
Branchless algorithms present a constant execution time, regardless of the value of the threshold, while  $if$-based approaches offer better performances as the value of $th$ increases. 
With $th \geq 0.3$, ``Vectorized IF'' is the most efficient approach, with a speedup up to $3\times$ compared to its na\"ive counterpart.

\vspace{1mm}
\noindent \textbf{Fast set membership}.
Once $\mathtt{close}_i^{th}$ is computed, we have to efficiently compute Equation~\ref{eq:our_filter_th}. 
Here, given $I_P$ as a list of integers, we have to test if at least one of its members $I_P^j$ belongs to $\mathtt{close}^{th}_i$, with $i = 1, \dots, n_q$.
This can be efficiently done using \emph{bit vectors} for representing $\mathtt{close}_i^{th}$.
A bit vector maps a set of integers up to $N$ into an array of $N$ bits, where 
the $e$-th bit is set to one if and only if the integer $e$ belongs to the set.
Adding and searching any integer $e$ can be performed in constant time with bit manipulation operators.
Moreover, bit vectors require $N$ bits to be stored. In our case,  since we have $|C| = 2^{18}$, a bit vector only requires  $32K$ bytes to be stored.

\begin{figure}[!t]
  \centering
  \begin{minipage}[b]{0.4\textwidth}
    \includegraphics[width=\textwidth]{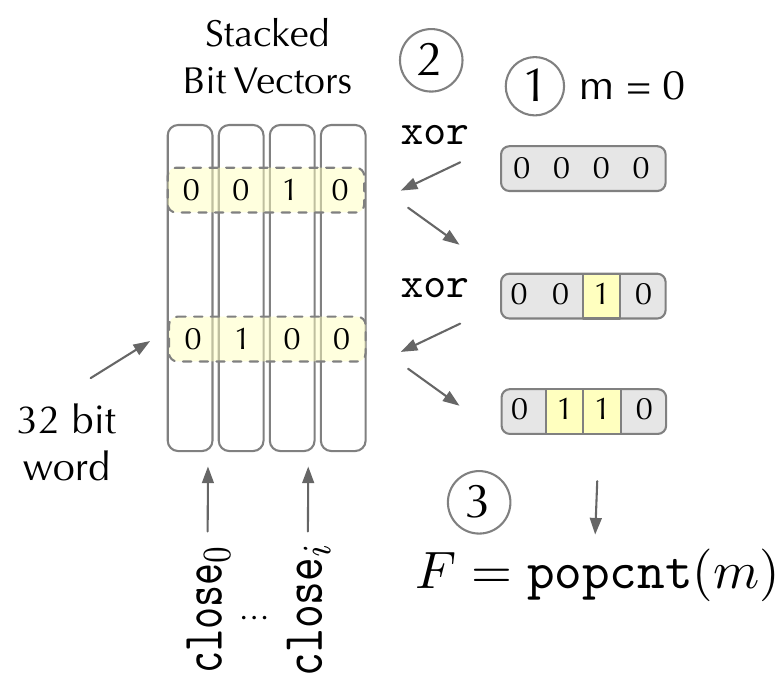}
    \caption{Vectorized Fast Set Membership algorithm based on bit vectors. \label{fig:stackedbitvect}}
  \end{minipage}
  \hfill
  \begin{minipage}[b]{0.5\textwidth}
  \includegraphics[width=\textwidth]{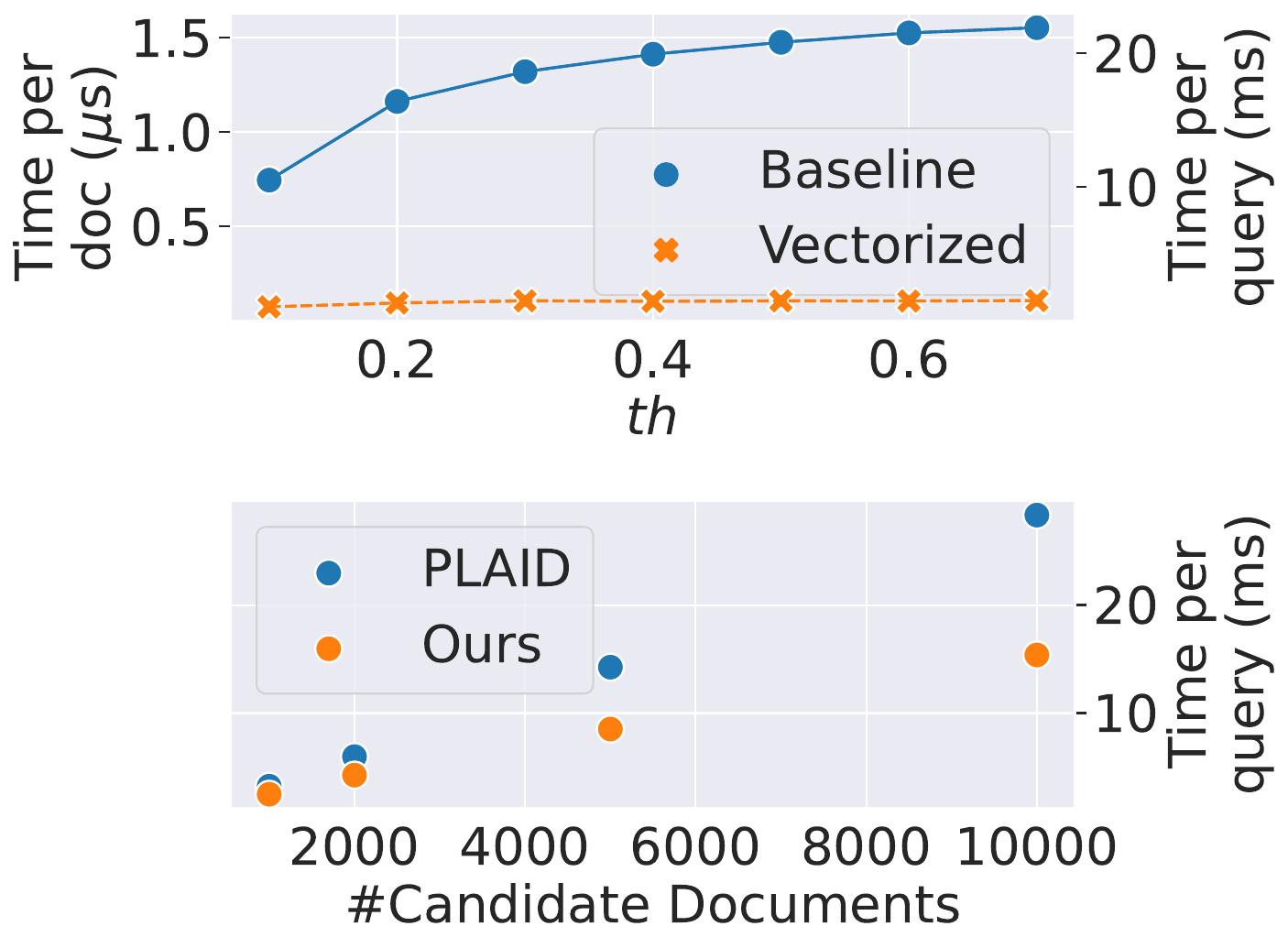}
    \caption{Vectorized vs na\"ive Fast Set Membership (\textbf{up)}. Ours vs PLAID filtering (\textbf{down}).\label{fig:mixed}}
  \end{minipage}
\end{figure}

Since we search through all the $n_q$ bit vectors at a time, we can further exploit the bit vector representation by stacking the bit vectors vertically (Figure~\ref{fig:stackedbitvect}). This allows to search a centroid index through all the $\mathtt{close}_i^{th}$ at a time. 
The bits corresponding to the same centroid for different query terms are consecutive and fit a $32$-bit word. This way, we can simultaneously test the membership for all the queries in constant time with a single bitwise operation.
In detail, our algorithm works by initializing a mask $m$ of $n_q=32$ bits at zeros (Step 1, Figure~\ref{fig:stackedbitvect}). Then, for each term in the candidate documents, it performs a bitwise \texttt{xor} between the mask and the $32$-bit word representing the membership to all the query terms (Step 2, Figure~\ref{fig:stackedbitvect}). 
Hence, Equation~\ref{eq:our_filter_th} can be obtained by counting the number of $1$s in $m$ at the end of the execution with the \texttt{popcnt} operation featured by modern CPUs (Step 3, Figure~\ref{fig:stackedbitvect}).

Figure~\ref{fig:mixed} (up) shows that our ``Vectorized'' set membership implementation delivers a speedup ranging from $10\times$ to $16\times$ a ``Baseline'' relying on a na\"ive usage of bit vectors. In particular, our bit vector-based pre-filtering can be up to $30\times$ faster than the centroid-interaction proposed in PLAID~\cite{santhanam2022plaid}, cf. Figure~\ref{fig:mixed} (down). 

\subsection{Fast Filtering of Candidate Passages} 
Our pre-filtering approach allows us to efficiently filter out non-relevant passages and is employed upstream of PLAID's centroid interaction (Equation~\ref{eq:plaid_approx}). 
We now show how to improve the efficiency of the centroid interaction itself.

Consider a passage $P$ and its associated centroid scores matrix  $\tilde{P} = q_i \cdot \bar{C}^{T_j}$. 
Explicitly building this matrix allows to reuse it in the scoring phase, in place of the costly decompression step (Section~\ref{subsec:docscoring}).
To build $\tilde{P}$, we transpose $CS$  into $CS^{T}$ of size $|C| \times n_q$.
The $i$-th row of $CS^{T}$ allows access to all the $n_q$ query terms scores for the $i$-th centroids. 
Given the ids of the closest centroids for each passage term (defined as $I_P$ in Section~\ref{subsec:docselection}) 
we retrieve the scores for each centroid id. We build $\tilde{P}^T$---$\tilde{P}$ transposed---to allow the CPU to read and write contiguous memory locations. This grants more than $2\times$ speedup compared to processing $\tilde{P}$. 
We now have $\tilde{P}^T$ of shape $n_t \times n_q$. We need to max-reduce along the columns and then sum the obtained values to implement Equation~\ref{eq:plaid_approx}.
This is done by iterating on the $\tilde{P}^T$ rows and packing them into AVX512 registers. Given that $n_q =32$, each AVX512 register can contain $512/32 = 16$ floating point values, so we need $2$ registers for each row. We pack the first row into \emph{max\_l} and \emph{max\_h}. All the successive rows are packed into \emph{current\_l} and \emph{current\_h}. At each iteration, we compare \emph{max\_l} with \emph{current\_l} and \emph{max\_h} with \emph{current\_h} using the \texttt{\_mm512\_cmp\_ps\_mask} AVX512 instruction described before. The output mask $m$ is used to update the \emph{max\_l} and \emph{max\_h} by employing the \texttt{\_mm512\_mask\_blend\_ps} instruction. The \texttt{\_mm512\_cmp\_ps\_mask} has throughput $2$ on IceLake Xeon CPUs, so  each row of $\tilde{P}$ is compared with \emph{max\_l} and \emph{max\_h} in the same clock cycle, on two different ports. The same holds for the \texttt{\_mm512\_mask\_blend\_ps} instruction, entailing that the max-reduce operation happens in $2$ clock cycles without considering the memory loading. 
Finally, \emph{max\_l} and \emph{max\_h} are summed together, and the function \texttt{\_mm512\_reduce\_add\_ps} is used to ultimate the computation. 

We implement PLAID's centroid interaction in C++ and we compare its filtering time against our SIMD-based solution. The results of the comparison are reported for different values of candidate documents in Figure~\ref{fig:mixed} (down). Thanks to the proficient read-write pattern and the highly efficient column-wise max-reduction, our method can be up to $1.8\times$ faster than the filtering proposed in PLAID. 
 


\subsection{Late Interaction}
\label{subsec:docscoring}
The $b$-bit residual compressor proposed in previous approaches~\cite{santhanam2022colbertv2,santhanam2022plaid} requires a costly decompression step before the late interaction phase.
Figure~\ref{fig:plaid_breakdown} shows that in PLAID
decompressing the vectors costs up to $5\times$ the late interaction phase.

We propose compressing the residual $r$ by employing Product Quantization (PQ)~\cite{jegou2010product}. PQ allows the computation of the dot product between an input query vector $q$ and the compressed residual $r_{pq}$ without decompression.
Consider a query $q$ and a candidate passage $P$. We decompose the computation of the max similarity operator (Equation~\ref{eq:maxsim}) into
\begin{equation}
\label{eq:ourmaxsim}
	S_{q, P} = \sum_{i=1}^{n_q} \max_{j =1\dots n_t} (q_i \cdot \bar{C}^{T_j} + q_i \cdot r^{{T_j}}) \simeq \sum_{i=1}^{n_q} \max_{j = 1\dots n_t} (q_i \cdot \bar{C}^{T_j} + q_i \cdot r_{pq}^{{T_j}}),
\end{equation}
where and $r^{{T_j}} = T_j - \bar{C}^{T_j}$. On the one hand, this decomposition allows to exploit the pre-computed $\tilde{P}$ matrix. On the other hand, thanks to PQ, it computes the dot product between the query and the residuals without decompression.

We replace PLAID's residual compression with PQ, particularly with JMPQ ~\cite{fang2022joint}, which optimizes the codes of product quantization during the fine-tuning of the language model for the retrieval task. We tested $m = \{16, 32\}$, where $m$ is the number of sub-spaces used to partition the vectors~\cite{jegou2010product}. We experimentally verify that PQ reduces the latency of the late interaction phase up to $3.6\times$ compared to PLAID $b$-bit compressor. Moreover, it delivers the same ($m=16$) or superior performance ($m=32)$ in terms of MRR@10 when leveraging the JMPQ version.

We propose to further improve the efficiency of the scoring phase by hinging on the properties of Equation~\ref{eq:ourmaxsim}. We experimentally observe that, in many cases, 
$q_i \cdot \bar{C}^T_j > q_i \cdot r_{pq}^{{T_j}}$, meaning that the \emph{max} operator on $j$, in many cases, is lead by the score between the query term and the centroid, rather than the score between the query term and the residual. We argue that it is possible to compute the scores on the residuals only for a reduced set of document terms $\bar{J}_i$, where $i$ identifies the index of the query term. In particular, $\bar{J}_i = \{j | q_i \cdot \bar{C}^T_j > th_{r} \}$, where $th_{r}$ is a second threshold that determines whether the score with the centroid is sufficiently large.
With the introduction of this new per-term filter, Equation~\ref{eq:ourmaxsim} now becomes computing the max operator on the set of passages in $\bar{J}_i$, i.e., 
\begin{equation}
\label{eq:ourmaxsimapprox}
	S_{q, P} = \sum_{i=1}^{n_q} \max_{j \in \bar{J}_i} (q_i \cdot \bar{C}^{T_j} + q_i \cdot r_{pq}^{{T_j}}).
\end{equation}

\begin{figure}[tb]
\centering
\includegraphics[width=\columnwidth]{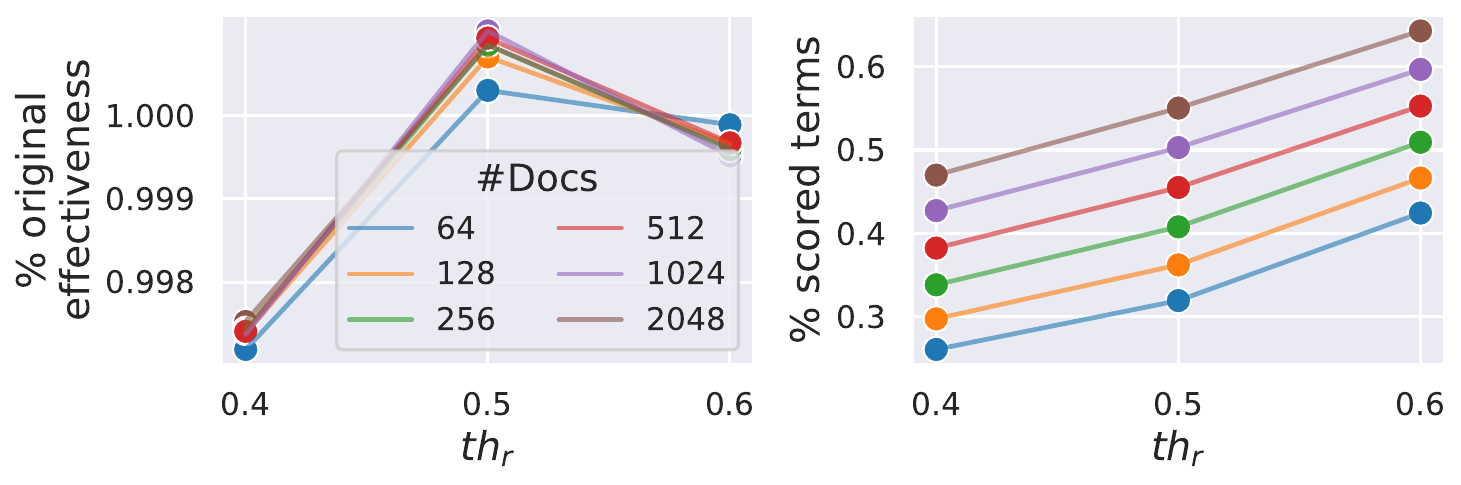}
\caption{Performance of our dynamic term-selection filtering for different values of $th_r$, in terms of percentage of original effectiveness (\textbf{left}) and in terms of percentage of original number of scored terms (\textbf{right}). The percentage of original effectiveness is computed as the ratio between the MRR@10 computed with Equation~\ref{eq:ourmaxsimapprox} and Equation~\ref{eq:ourmaxsim}.\label{fig:document_term_filtering}}
\end{figure}

In practice, we compute the residual scores only for those document terms whose centroid score is large enough. If $\bar{J}_i = \emptyset$, we compute $S_{q, P} $ as in Equation~\ref{eq:ourmaxsim}. 
Figure~\ref{fig:document_term_filtering} (left) reports the effectiveness of our approach. On the $y$-axis, we report the percentage of the original effectiveness, computed as the ratio between the MRR@10 computed with Equation~\ref{eq:ourmaxsimapprox} and Equation~\ref{eq:ourmaxsim}. Filtering document terms according to Equation~\ref{eq:ourmaxsimapprox} does not harm the retrieval quality, as it delivers substantially the same MRR@10 of Equation~\ref{eq:ourmaxsim}.
On the right side of Figure~\ref{fig:document_term_filtering}, we report the percentage of scored terms compared to the number of document terms computed using Equation~\ref{eq:ourmaxsim}. With $th_r = 0.5$, we are able to reduce the number of scored terms of at least $30\%$ (right) without any performance degradation in terms of MRR@10.

\section{Experimental Evaluation}
\label{sec:exp}

\noindent \textbf{Experimental Settings}. This section compares our methodology against the state-of-the-art engine for multi-vector dense retrieval, namely PLAID~\cite{santhanam2022plaid}. We conduct experiments on the \msmarco passages dataset~\cite{nguyenms} for the in-domain evaluation and on LoTTE~\cite{santhanam2022colbertv2} for the out-of-domain evaluation. We generate the embeddings for \msmarco using the \colberttwo model. The generated dataset is composed of about $600$M $d$-dimensional vectors, with $d = 128$.
Product Quantization is implemented using the FAISS \cite{DBLP:journals/corr/JohnsonDJ17} library, and optimized using the JMPQ technique~\cite{fang2022joint} on \msmarco.
The implementation of EMBV is available on Github\footnote{\url{https://github.com/CosimoRulli/emvb}}. We compare EMVB against the original PLAID implementation~\cite{santhanam2022plaid}, which also implements its core components in C++.
Experiments are conducted on an Intel Xeon Gold 5318Y CPU clocked at $2.10$ GHz, equipped with the AVX512 instruction set, with single-thread execution. Code is compiled using GCC 11.3.0 (with \texttt{-O3} compilation options) on a Linux 5.15.0-72 machine.
When running experiments with AVX512 instruction on 512-bit registers, we ensure not to incur in the frequency scaling down event reported for Intel CPUs~\cite{lemire-avx}.



\smallskip
\noindent \textbf{Evaluation}.
Table~\ref{tab:cmpplaid} compares EMVB against PLAID on the \msmarco dataset, in terms of memory requirements (num. of bytes per embedding), average query latency (in milliseconds), MRR@$10$, and Recall@$100$, and $1000$.

\begin{table}[!htb]
\centering
\begin{tabular}{llrrrrr} 
\toprule
$k$ & Method & Latency  ($msec$.) & Bytes & MRR@10 & R@100 & R@1000\\
\midrule
\multirow{3}{*}{10} & PLAID &  $131$ \textcolor{white}{$(1.0\times)$} & 36 & 39.4 & - & - \\
& EMVB (m=16) & $62$ $(2.1\times)$ & 20 & 39.4 & - & - \\
& EMVB (m=32) & $61$ $(2.1\times)$ & 36 & 39.7 & - & - \\
\midrule
\multirow{3}{*}{100} & PLAID &  $180$ \textcolor{white}{$(1.0 \times)$} & 36 & 39.8 & 90.6 & -\\
& EMVB (m=16) & $68$ $(2.6\times)$ & 20 & 39.5 & 90.7 & - \\
& EMVB (m=32) & $80$ $(2.3\times)$ & 36 & 39.9 & 90.7 & - \\
\midrule 
\multirow{3}{*}{1000} & PLAID & $260$ \textcolor{white}{$(1.0\times)$} & 36 & 39.8 & 91.3 & 97.5 \\
& EMVB (m=16) & $93$ $(2.8\times)$ & 20 & 39.5 & 91.4 & 97.5 \\
& EMVB (m=32) & $104$ $(2.5\times)$ & 36 & 39.9 & 91.4 & 97.5 \\
\bottomrule
\end{tabular}
\caption{Comparison between EMVB and PLAID in terms of average query latency, number of bytes per vector embeddings, MRR, and Recall on \msmarco.\label{tab:cmpplaid}}
\end{table}

Results show that EMVB delivers superior performance along both the evaluated trade-offs. With $m=16$, EMVB almost halves the per-vector memory burden compared to PLAID, while being up to $2.8\times$ faster with almost no performance degradation regarding retrieval effectiveness. By doubling the number of sub-partitions per vector, i.e., $m=32$, EMVB outperforms the performance of PLAID in terms of MRR and Recall with the same memory footprint with up to $2.5\times$ speed up.

Table~\ref{tab:cmlotte} compares EMVB and PLAID in the out-of-domain evaluation on the LoTTE dataset. As in PLAID~\cite{santhanam2022plaid}, we employ Success@5 and Success@100 as retrieval quality metrics. On this dataset, EMVB offers slightly inferior performance in terms of retrieval quality. 
Recall that JMPQ~\cite{fang2022joint} cannot be applied in the out-of-domain evaluation due to the lack of training queries. Instead, we employ Optimized Product Quantization (OPQ)~\cite{ge2013optimized}, which searches for an optimal rotation of the dataset vectors to reduce the quality degradation that comes with PQ.
To mitigate the retrieval quality loss, we only experiment PQ with $m=32$, given that an increased number of partitions offers a better representation of the original vector.
On the other hand, EMVB can offer up to $2.9\times$ speedup compared to PLAID. This larger speedup compared to \msmarco is due to the larger average document lengths in LoTTE. In this context, 
filtering nonrelevant documents using our bit vector-based approach has a remarkable impact on efficiency. Observe that for the out-of-domain evaluation, our prefiltering method could be ingested into PLAID. This would allow to maintain the PLAID accuracy together with EMVB efficiency. Combinations of PLAID and EMVB are left for future work. 

\begin{table}[htb]
\centering
\begin{tabular}{llrrrr} 
\toprule
$k$ & Method & Latency ($msec$.) & Bytes & Success@5 & Success@100 \\
\midrule
\multirow{2}{*}{10}& PLAID &  $131$ \textcolor{white}{$(1.0 \times)$} & 36 & 69.1 & -\\
& EMVB (m=32)& $82$ $(1.6\times)$ &36& 69.0 & - \\
\midrule 
\multirow{2}{*}{100}& PLAID &  $202$ \textcolor{white}{$(1.0 \times)$} & 36 & 69.4 & 89.9\\
& EMVB (m=32) & $129$ $(1.6\times)$ & 36 & 69.0 & 89.9 \\
\midrule
\multirow{2}{*}{1000}& PLAID &  $411$ \textcolor{white}{$(1.0 \times)$} & 36 & 69.6 & 90.5\\
& EMVB (m=32)& $142$ $(2.9\times)$ & 36 & 69.0 & 90.1\\
\bottomrule
\end{tabular}
\caption{Comparison between EMVB and PLAID in terms of average query latency, number of bytes per vector embeddings, Success@5, and Success@100 on LoTTE.\label{tab:cmlotte}}
\end{table}

\section{Conclusion}
\label{sec:conclusion}
We presented EMVB, a novel framework for efficient multi-vector dense retrieval. EMVB advances PLAID, the current state-of-the-art approach, by introducing four novel contributions. First, EMVB employs a highly efficient pre-filtering step of passages using optimized bit vectors for speeding up the candidate passage filtering phase. Second, the computation of the centroid interaction is carried out with reduced precision. Third, EMVB leverages Product Quantization to reduce the memory footprint of storing vector representations while jointly allowing for fast late interaction. Fourth,  we introduce a per-passage term filter for late interaction, thus reducing the cost of this step of up to $30\%$.
We experimentally evaluate EMVB against PLAID on two publicly available datasets, i.e., \msmarco and LoTTE. Results show that, in the in-domain evaluation, EMVB is up to $2.8\times$ faster, and it reduces by $1.8\times$ the memory footprint with no loss in retrieval quality compared to PLAID. In the out-of-domain evaluation, EMVB is up to $2.9\times$ faster with little or no retrieval quality degradation.

\smallskip
\noindent \textbf{Acknowledgements}.
This work was partially supported by the EU - NGEU, by the PNRR - M4C2 - Investimento 1.3, Partenariato Esteso PE00000013 - ``FAIR - Future Artificial Intelligence Research'' - Spoke 1 ``Human-centered AI'' funded by the European Commission under the NextGeneration EU program, by the PNRR ECS00000017 Tuscany Health Ecosystem Spoke 6 ``Precision medicine \& personalized healthcare'', by the European Commission under the NextGeneration EU programme, by the Horizon Europe RIA ``Extreme Food Risk Analytics'' (EFRA), grant agreement n. 101093026, by the ``Algorithms, Data Structures and Combinatorics for Machine Learning'' (MIUR-PRIN 2017), and by the ``Algorithmic Problems and Machine Learning'' (MIUR-PRIN 2022).

\bibliographystyle{splncs04}
\bibliography{biblio}
\end{document}